\documentclass{aa}
\usepackage{natbib,psfig}
\def\EEL{\textit{Energy--Entropy line}{}}
\def\Se{S\'ersic{}}
\def\SE{specific entropy{}}
\def\EM{energy--mass{}}
\def\DV{de Vaucouleurs{}}
\def\diff{{\rm d}}
\def\eff{{\rm eff}}

\begin{document}

% \thesaurus{02.07.1; 03.13.8; 11.03.4;
% 11.05.1; 11.05.2; 11.06.1; 11.06.2; 11.19.6}

\title{Energy, entropy and mass scaling relations for elliptical galaxies. 
Towards a physical understanding of their photometric properties}

\author{I. M\'arquez\inst{1} \and G.B. Lima Neto\inst{2} \and H. Capelato\inst{3}
\and F. Durret\inst{4,5} \and B. Lanzoni\inst{4,6} \and D. Gerbal\inst{4,5}}

\offprints{I. M\'arquez (isabel@iaa.es)}

\institute{
Instituto de Astrof\'\i sica de Andaluc\'\i a (C.S.I.C.), Apartado 3004 , 
E-18080 Granada, Spain 
\and
Instituto Astron\^omico e Geof\'{\i}sico/USP, Av. Miguel Stefano 4200, 
S\~ao Paulo/SP, Brazil 
\and
Instituto Nacional de Pesquisas Espaciais, S\~ao Jos\'e dos Campos/SP, 
Brazil 
\and
Institut d'Astrophysique de Paris, CNRS, 98bis Bd Arago, F-75014 Paris, 
France 
\and
DAEC, Observatoire de Paris, Universit\'e Paris VII, CNRS (UA 173), 
F-92195 Meudon Cedex, France 
\and
Osservatorio Astronomico di Bologna, via Ranzani 1, 40127 Bologna, Italy
}

\date{Accepted ????. Received ????; Draft printed: \today}

\authorrunning{M\'arquez et al.}
\titlerunning{Energy, entropy and mass scaling relations for Es}

\abstract{
In the present paper, we show that elliptical galaxies (Es) obey a
scaling relation between potential energy and mass. Since they are relaxed
systems in a post violent-relaxation stage, they are quasi-equilibrium
gravitational systems and therefore they also have a quasi-constant specific
entropy. Assuming that light traces mass, these two laws imply that in the
space defined by the three S\'ersic law parameters (intensity $\Sigma_0$,
scale $a$ and shape $\nu$), elliptical galaxies are distributed on two
intersecting 2-manifolds: the \textit{Entropic Surface} and the
\textit{Energy--Mass Surface}. Using a sample of 132 galaxies belonging to
three nearby clusters, we have verified that ellipticals indeed follow these
laws.
This also implies that they are distributed along the intersection line (the
\textit{Energy--Entropy line}), thus they constitute a one-parameter family.
These two physical laws (separately or combined), allow to find the
theoretical origin of several observed photometrical relations, such as the
correlation between absolute magnitude and effective surface brightness, and
the fact that ellipticals are located on a surface in the $[\log R_\eff, -2.5
\log \Sigma_0, \log \nu]$ space. 
The fact that elliptical galaxies are a one-parameter family has important
implications for cosmology and galaxy formation and evolution models.
Moreover, the \textit{Energy--Entropy line} could be used as a 
distance indicator. 
\keywords{galaxies: clusters: individual (Coma, Abell 85,
Abell 496) --- galaxies: elliptical and lenticular, cD --- galaxies:
fundamental parameters --- distance scale --- gravitation} 
} 

\maketitle

\section{Introduction}

Elliptical galaxies (Es) present a striking regularity in their global
luminosity distributions in the sense that, within a wide range of sizes,
their light profiles can be described by simple functions, such as the
S\'ersic law \citep{Caon, Graham, Prugniel}, which is a generalization of the
de Vaucouleurs $R^{1/4}$ profile. This regularity implies that ellipticals
constitute a well defined family of galaxies. The regular properties of Es
have been the subject of different approaches, concerning both photometric and
spectroscopic parameters, resulting in well known relations such as the
Faber-Jackson and Kormendy relations \citep{FaberJackson, Kormendy} or the
Fundamental Plane \citep{Djorgovski, Dressler}. Es are supposed to be formed
under non collisional processes, where dissipation is expected to be
negligible. Under these circumstances, both the initial conditions and the
gravitational forces are expected to have a crucial influence on their
properties.

In previous papers, we have addressed the question of how the properties of
Es may be a consequence of the physical laws they should obey. In
\citet{Gerbal} and \citet[hereafter Paper I]{Lima}, we have calculated the
specific entropy of Es and shown that the results of analysing the 
cluster Es with S\'ersic profiles are compatible with a \textit{unique
value} for this specific entropy, assuming a constant mass/luminosity ratio.
In a subsequent paper, \citet[hereafter Paper II]{Marquez}, we have shown
that the specific entropy of Es was in fact correlated with the logarithm
of the total luminosity while the specific entropy of simulated halos of
collisionless particles was correlated with the logarithm of the total
mass. Moreover, we suggested that another physical law must be operating in
order to explain that Es reside in a thin line embedded in a surface
defined by a (almost) constant specific entropy.

In this paper, we show that this second physical law is in fact a
scaling relation between the potential energy and mass; photometrical
relations can be naturally derived from this scaling law under some
simplifying hypotheses. We also analyse what the consequences on the
origin and evolution of Es are. In Sect. 2 we present a short
theoretical introduction to the concepts we will deal with, from which
we derive a possible explanation for the existence of an energy--mass
relation, and we describe the passage from theoretical parameters to
observational ones using the S\'ersic profile. We improve our analysis
taking into account a possible dependence of the mass to light ratio
on the total luminosity. In Sect. 3 we show that similar scaling
relations are also found in other types of self-gravitating systems.
In Sect. 4 we obtain the relations among the S\'ersic parameters and
we compare the theoretical predictions with the results from
observational data in Sect. 5. The discussion and conclusions are
given in Sect. 6.

\section{Theoretical background}\label{sec:deuxlois}

In the standard model of structure formation, initially small
perturbations in the cosmic density field start growing in amplitude
at the same rate. At a given epoch, when their overdensity relative to
the average density is larger than a critical threshold that depends
on the assumed cosmology, they stop expanding with the Universe and
collapse. Bound objects then settle in an equilibrium configuration
through gravitational and radiative processes.

In the following (Sect. \ref{subsec:equilibre} and \ref{subsec:scaling}), we
propose two laws that elliptical galaxies should obey if they form and reach
(quasi) equilibrium under the action of gravitational processes only. As a
first approximation, we will assume that radiative processes play a minor role
compared to gravitation.

\subsection{Specific entropy}\label{subsec:equilibre}

Elliptical galaxies are thought to be in a quasi-equilibrium state, implying
that they should obey the virial theorem. The second law of thermodynamics
states that a system in equilibrium is in a maximum entropy configuration.
Since elliptical galaxies are gravitational systems, they never really reach
an equilibrium state during the time scale of the two-body relaxation.
However, even if the entropy $S$ of an E galaxy is ever increasing on a
secular time scale, after violent relaxation one may consider that the system
is in a quasi-equilibrium stage, which is equivalent to saying that the
entropy is quasi-stationary.

Several works have been devoted to the problem of the entropy in gravitational
systems (see for instance \citet{Merritt}, and references therein). In a
previous paper (Paper~I), we have shown that, assuming that an elliptical
galaxy is in a quasi-equilibrium stage, spherically symmetric with an
isotropic velocity distribution and constant mass--luminosity ratio, a
thermodynamic entropy function can be defined.

In Paper II, instead of using the thermodynamic definition of the entropy,
we adopted the microscopic Boltzmann-Gibbs entropy, defined from the
distribution function in the 6-dimensional phase space. In order to compare
objects of different masses in a consistent way, we have introduced the
specific entropy, i.e. the entropy normalized to the mass:
\begin{equation}
    s \equiv \frac{S}{M} = \frac{- \int f \ln f \diff^{3}x \diff^{3}v}%
    {\int f \diff^{3}x \diff^{3}v} \, ,
\end{equation}
where $f$ is the distribution function. We have then shown that the specific
entropy $s$, computed under the previous hypotheses and using the S\'ersic
profile is almost the same among Es (within 10\%). In fact, based on $N$-body
simulations, we suggested that $s$ is not strictly unique but, as a result of
merging processes, weakly depends on the logarithm of the galactic mass. We
then showed that the specific entropy and the luminosity are tightly
correlated. Adopting an appropriate $M/L$ ratio (see Sect.~\ref{sec:simplify}
below) this translates to:
\begin{equation}
  s=s_{0} + \delta_{s} \ln M \, ,
\label{eq:entro}
\end{equation}
where $s_{0}$ and $\delta_{s}$ are to be determined by observations.

\subsection{Scaling relation between potential energy and
mass}\label{subsec:scaling}

Let us consider a region with a characteristic length $\lambda_{i}$ and a mean
density $\rho_{i}$, in the initial density field (all quantities will be indexed
with $i$ -- for initial). This region is supposed to be the location of a
present E galaxy (its properties will be indexed with $p$ -- for present).

One may suppose, at least to zeroth order, that the following quantities are
conserved during the formation and subsequent evolution of the structure:
\begin{enumerate}
     \item the total mass: $\quad M_{i}=M_{p}=M$;
     \item the total energy: $\quad E_{i}=E_{p}=E$. 
\end{enumerate}
The total energy at the initial time is:
\begin{equation}
     E =T_{i} + U_{i} \simeq U_{i} \, ,
\label{eq:energinitiale}
\end{equation}
(where $T$ and $ U$ are the kinetic and potential energies), because
the kinetic energy is negligible compared to the potential energy.

At the present stage, when the galaxy is already relaxed, the two following
relations are verified:
\begin{equation}
    T_{p} + U_{p}=E \, , 
\label{eq:energie}
\end{equation}
and
\begin{equation}
    2T_{p} + U_{p} = 0 \, .
\label{eq:viriel}
\end{equation}
Eq.~(\ref{eq:viriel}) is the virial relation, which of course does not hold
in the $i$-case.

Playing the game of addition, subtraction and comparison, we finally obtain the
usual relation for an initially cold gravitational system that collapses and
reaches equilibrium:
\begin{equation} 
    U_{p} \simeq 2 U_{i} \, .
\label{eq:approx}
\end{equation}
Notice that, if the initial kinetic energy is taken into account, the above 
relation should read $U_{p} = 2 U_{i} + 2 T_{i}$.

In the initial linear regime all quantities, such as the typical
perturbation length $\lambda_{i}$, its mass $M$, and its potential energy
$U$, are related to one another by scaling relations, e.g. (we will adopt
hereafter $G=1$):
\begin{equation}
    M \propto \lambda_{i}^3 \quad \mbox{and} \quad U_{i}\propto
M^2/\lambda_{i} \, ,
\end{equation}
implying that:
\begin{equation} 
    U_{i}\propto \frac{M^2}{M^{1/3}}\propto M^{5/3}.
\label{eq:relechelle}
\end{equation}
Using relation~(\ref{eq:approx}), we then obtain:
\begin{equation} 
    U_{p} \propto M^{5/3} \, .
\label{eq:ouf}
\end{equation}
If we consider the (small) contribution of the initial kinetic energy then the
above relation will be tilted, $U_{p} \propto M^{\varepsilon + 5/3}$, where
$\varepsilon$ is a small number depending on the exact initial conditions,
e.g., the contribution of $T_{i}$ to the total energy and the initial density
fluctuation power spectrum.

Equation (\ref{eq:ouf}) can be also written as:
\begin{equation} 
    \ln(U_{p}) -5/3\ln(M) = e_0 \, ,
\label{eq:econst}
\end{equation}
where $e_0$ is a constant to be determined by observations.

\subsection{From theory to observations}\label{subsec:remarques}

\subsubsection{Simplifying Hypotheses}\label{sec:simplify}

The relations given in Eqs.~(\ref{eq:entro}) and (\ref{eq:econst}) are both
theoretically motivated laws that need to be translated into observable
quantities if one wants to test them observationally. In other words, we 
need to go from the observed two dimensional light distribution to a three 
dimensional mass distribution. Rather than doing a detailed dynamical and 
morphological model, we will assume some simplifying hypotheses: spherical 
symmetry, isotropic velocity dispersion tensor and a mass to light ratio 
independent of galactic radius $R$.

The last of these hypotheses, i.e., $M/L \equiv \mathcal{F}(R) =
\mbox{constant}$, implies that the mass distribution follows the
light profile, with the same scale length and shape. Notice that
the above ``constant'' may in fact depend on the total galaxy mass or
luminosity (as we will see below). This hypothesis is clearly a rough
simplification: X-ray observation of giant ellipticals
\citep[e.g.][]{Loewenstein}, gravitational lensing analyses
\cite[e.g.][]{Griffiths}, and dynamical studies
\cite[e.g.][]{Gerhard}, suggest that $\mathcal{F}(R)$ increases with
radius. This increase, however, seems to be important only beyond $R \ga 2
R_{\eff}$ and current data is compatible with $\mathcal{F}(R) \approx$ 
constant for $R \la 2 R_{\eff}$ \citep{Kronawitter}.

It has been argued \citep{Bertin} that a significant amount of dark
matter can coexist with the stellar component without distorting the
luminosity profile either if both the stellar and dark matter profiles
have the same shape or if the dark matter is much more diffuse, i.e.,
with a larger scale length and lower central density. The former case
corresponds to our hypothesis of a constant $\mathcal{F}(R)$. In the
latter case, the contribution of the dark matter component will be
important only in the external region of the galaxy.

The spherical symmetry and isotropy hypotheses imply that the
distribution function is simply a function of one variable, the
binding energy. Although these hypotheses may hold in the internal
region, they must be poor approximations in the external regions of
the most flattened galaxies. However, there is no clear correlation
between ellipticity and luminosity and the effect of neglecting the
flattening should not produce any systematic error.

Finally, we will assume that mass and luminosity are related by a power-law:
\begin{equation}
    \mathcal{F}(L) = k_{F} L^{\alpha} \, ,
    \label{eq:M/Lfonction}
\end{equation}
where the index $\alpha$ may be zero (equal mass to light ratio for all
ellipticals). The above functional form is suggested, e.g., by the analysis of
the tilt of the Fundamental Plane \textit{assuming homology}
\citep[e.g.,][]{Burstein, Jorgensen}, where $\alpha \approx 0.22$--0.32.
Notice that the hierarchical merger simulations discussed by \cite{Capelato,
Capelato2} give a natural explanation for the origin of the Fundamental Plane
under the hypothesis of a constant value of $\mathcal{F}$.

\subsubsection{Working model}

With the above hypotheses, the mass, potential energy and specific
entropy are easy to compute from the light distribution.

Our working model is the S\'ersic \citeyearpar{Sersic} luminosity profile:
\begin{equation} \Sigma = \Sigma_0 \exp[-(R/a)^{\nu}] \, ,
     \label{def}
\end{equation}
which describes well the observed surface profile distribution of ellipticals
\citep[e.g.,][see also Sect. \ref{subsec:discussions}]{Caon, Graham,
Prugniel}, and allows to compute analytically good approximations of all the
needed quantities. We therefore adopt this profile to model the observed
surface brightness distribution of a sample of elliptical galaxies (see
Sect.~\ref{subsec:nudist}), and to compute the physical quantities of
interest, like the total mass, the potential energy, the specific entropy,
etc. (see Tables~\ref{tbl:formule2} and \ref{tbl:formule1} in the Appendix for
a summary of the formulae we use).

Note that the S\'ersic law is a non-homologous generalization of the \DV\
profile, in the sense that a third parameter describing the shape of the
distribution (the structural parameter $\nu$) is introduced and left free,
instead of being fixed to 0.25. This of course allows a better fit to the
observations \citep[and Papers I and II]{Graham}.

We assume that only one dynamical component describes an elliptical galaxy and
ignore the possible presence of nuclear cusps or embedded discs. This
simplification is motivated by the fact that these components, if present,
usually represent at most a few percent of the total galactic mass
\citep[c.f.,][ and references therein; see also \cite{Saglia}]{deZeeuw,
Kormendy2}.
 
We stress here that we are doing a pure photometrical analysis and therefore
such a working model is required (with all its simplifying hypotheses) so that
we are able to derive dynamically related quantities like the specific entropy
or the potential energy.

\section{Scaling relations: Elliptical galaxies only?}\label{halos}

The derivation of the scaling relation between potential energy and
mass is based on fundamental principles (mass and energy
conservation). It can therefore be applied to any object formed by the
growth of density perturbations, driven by gravitational
interaction. Dark matter (DM) halos are therefore expected to follow
the same relation.

Assuming that collisionless $N$-body simulations provide a suitable
description of DM halo formation, relation~(\ref{eq:econst}) should then also
be found automatically in the results of such simulations. We have tested this
hypothesis by considering a cosmological $N$-body simulation with $512^3$
particles in a 479 Mpc$/h$ side box, characterized by the following
parameters: a $\Lambda$CMD cosmological model, with density parameters
$\Omega_0 = 0.3$ and $\Omega_\Lambda = 0.7$, Hubble constant
$H=100\,h \,$km~s$^{-1}$Mpc$^{-1}$, $h=0.7$, and normalization
$\sigma_8=0.9$ \citep{Yoshida}. The corresponding particle mass is
$6.8\times10^{10} h^{-1} {\rm M}_\odot$. A sphere of radius $r=7 h^{-1}$Mpc
around a very massive DM halo (M $\simeq 2.3\times 10^{15} h^{-1}\,{\rm
M}_\odot $, virial radius $R_{\rm vir} \simeq 2.7 h^{-1}$Mpc) has been
selected in the original simulation, and re--simulated at higher resolution (a
factor $\sim 35$ increase, to get a mass particle of about $2\times 10^9
h^{-1} \,{\rm M}_\odot$). Assuming a minimum mass of 10 particles per halo,
this allows to resolve halos down to $2\times 10^{10}h^{-1}\,{\rm M}_\odot$.

\begin{figure}[htb]
\centering
\mbox{\psfig{figure=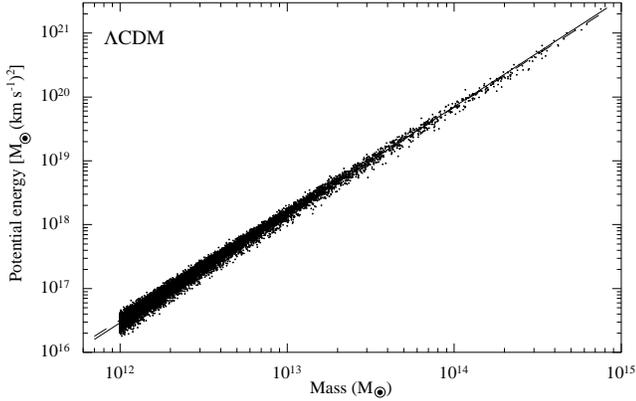,width=8.5cm}}
\caption[]{Potential energy--mass relation for about 29\,000 dark matter halos
selected in a high-resolution $N$-body simulation (see text). The solid line
is the best fit to the data, with a slope of $1.69\pm 0.02$, in very good
agreement with the expected theoretical value of 5/3 (dashed line).}
\label{fig:barbara}
\end{figure}

Fig.~\ref{fig:barbara} shows the distribution of potential energy
\textit{versus} mass for the halos found within the selected sphere, their
masses ranging from $10^{12}$ to $10^{15}h^{-1}\,{\rm M}_\odot$. We have
selected the halos that had their kinetic to potential energy ratio closer to
the virial theorem value, i.e., the ones satisfying $0.8 \le 2T/|U| \le 1.2$.
Due to the high resolution of the simulation, the computation of the potential
energy is very accurate, at least for all halos more massive than $\sim
10^{11} \mathrm{M}_{\odot}$, i.e. composed by more than 100 particles. 

A fit to the data indicates a power-law index of $1.69\pm 0.02$, very close to
the theoretical value of 5/3. The difference may be due to the fact that the 
$N$-body simulation takes into account all dynamical processes (mass and 
energy exchange and/or loss) self-consistently.

We have also verified that this relation is independent on cosmology. In fact,
we found the same slope of the potential-energy mass relation for DM halos in
all the GIF Simulations \citep{Kauffmann}, run for four different cosmological
models: SCDM, OCDM, $\Lambda$CDM, $\tau$CDM.

A comparable simulation of dark matter haloes was done by \cite{Jang-Condell}
in a much lower mass range. In this simulation, each particle represents $4.6
\times 10^4\ M_\odot$, and the range of halo masses is $4 \times 10^5\
M_\odot$ to $4 \times 10^8\ M_\odot$. The potential energy calculated for 2501
bound halos also seems to follow a power law $|U| \propto\ M^{5/3}$.

This is an example of the \textit{universality} of the scaling relation
between potential energy and mass. Whereas out of the scope of this paper,
numerous questions obviously arise, such as the dependence (or independence)
of the index with redshift or with the cosmology assumed in the simulation.

Interestingly, the analysis of ROSAT-PSPC images of 24 clusters has
also revealed the existence of a scaling law between the cluster
potential energy and the mass of the X-ray gas, with a dependence $U
\propto\ M^{1.72\pm 0.05}$, again close to 5/3 (see Durret et
al. 2001). Such a scaling law therefore appears to be general for
self-gravitating systems.

\section{Predicted relations among the \Se\ parameters}
\label{sec:Sersic}

We now show that the two relations introduced in Sects.~\ref{subsec:equilibre}
and \ref{subsec:scaling}, constraining the distribution of gravitational
matter in galaxies, define two surfaces intersecting each other in the \Se\
parameter space [$\nu, a, \Sigma_{0}$].

The first relation comes from the relation between the specific entropy, $s$,
and the mass, $M$, that can be expressed as
\begin{equation}
 s = s_{0} + \delta_{s} \ln M = s_{0} + \delta_{s} \ln (\mathcal{F} L) \, , 
    \label{eq:tiltEntro}
\end{equation}
cf. Eq.~(\ref{eq:entro})\footnote{In Paper~II, the specific entropy is defined
with a different normalization: $s' = -\frac{1}{M}\int \diff x^{3} \diff v^{3}
f \ln f$, where $\int \diff x^{3} \diff v^{3} f=M$. Here, we use the 
normalization $\int \diff x^{3} \diff v^{3} f^{*} = 1$ and compute the 
specific entropy $s$ using the normalized distribution function 
$f^{*}$. With simple algebra it can be shown that $s = s' +
\ln M$.}. From the expression of the entropy in Table~\ref{tbl:formule2}, it
is straightforward to obtain the \textit{Entropic Surface} in terms of the
\Se\ parameters, Eq.~(\ref{eq:PlanEntroTilt}) shown in the Appendix.

The second relation is obtained using the definition of potential energy and
mass given in Tables~\ref{tbl:formule2} and \ref{tbl:formule1}, that 
can be cast in a general form:
\begin{equation}
    U = k_{U} M^{\beta} \; \mbox{or, equivalently, } \ln U = e_{0} + 
    \beta \ln (\mathcal{F} L) \, ,
    \label{eq:U-M}
\end{equation}
where $e_{0} \equiv \ln k_{U}$ and we have dropped the index ``$p$''.
Using the luminosity and the potential energy derived for a mass
distribution following the S\'ersic profile (cf.
Table~\ref{tbl:formule2}), it is easy to obtain the
\textit{Energy--Mass Surface} in terms of the observed photometric
S\'ersic parameters. The formula describing this surface is shown in
the Appendix, Eq.~(\ref{eq:PlanEnergMass}).

\begin{figure}[h]
\centering
\mbox{\psfig{figure=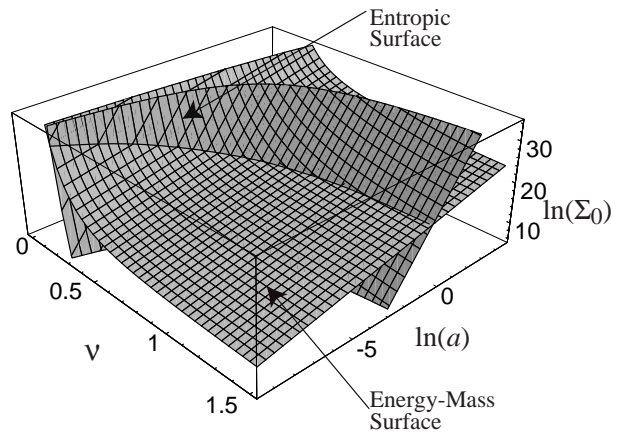,width=8cm}}
\caption[]{3-D representation of the \SE\ and
\EM\ 2-manifolds, using the coordinates: [$\log \Sigma_{0},\log a,
\nu$]. In this example, we assume $M \propto L$.}
\label{fig:intersection}
\end{figure}

A three dimensional representation of these two surfaces is given in
Fig.~\ref{fig:intersection}, in the [$\nu, a, \Sigma_{0}$] space. The
intersection line (called the Entropy--Energy line in Paper II) is the locus
on which elliptical galaxies are 
distributed in the space of the
S\'ersic parameters.

Combining relations (\ref{eq:PlanEntroTilt}) and (\ref{eq:PlanEnergMass}),
i.e. calculating the intersection of the Entropic and Energy--Mass surfaces,
two-by-two relations between the \Se\ parameters can be obtained. The general
[a, $\nu$] and [$\Sigma_{0}, \nu$] relations are given in the Appendix. For
the particular case where $\alpha = 0$ (constant $M/L$ ratio), $\beta = 5/3$
(cf. Sect—\ref{subsec:scaling}) and $\delta_{s} = 0$ (unique specific
entropy), we have for the [a, $\nu$] relation:
%-----------------------------------------------------------------
%
\begin{eqnarray}
\ln(a) + \mathrm{F}_{a}(\nu) &=& -\frac{3}{4} e_{0} +
       \frac{1}{6} s_{0} +\frac{3}{4}\ln G \, ,
      \label{eq:correlationsanu} \\
{\rm with}\ \mathrm{F}_{a}(\nu) &=& \frac{3}{4}\ln R^*_{g}(\nu) -
\frac{1}{4}\ln M_2^*(\nu) + \frac{1}{6} F_{2}(\nu) \, , \nonumber
\end{eqnarray}
and, for the [$\Sigma_{0}, \nu$] relation:
%
%-----------------------------------------------------------------
\begin{eqnarray}
\ln(\Sigma_{0}) + \mathrm{F}_{\Sigma}(\nu) &=& \frac{9}{4} e_{0} + 
                 \frac{1}{6} s_{0} - \frac{9}{4}\ln G \, ,
                 \label{eq:correlationsigmanu} \\
{\rm with}\ \mathrm{F}_{\Sigma}(\nu) &=& - \frac{9}{4}\ln R^*_{g}(\nu) +
\frac{3}{4}\ln M_2^*(\nu) + \frac{1}{6} F_{2}(\nu) \, . \nonumber
\end{eqnarray}
The relation between $a$ and $\Sigma_0$ is obtained by combining
relations~(\ref{eq:correlationsanu}) and (\ref{eq:correlationsigmanu}),
with $\nu$ as a parameter.

\section{Predictions faced to observations}\label{sec:confrontations}

\subsection{Data set and distribution of the $\nu$
parameter}\label{subsec:nudist}

The sample of elliptical galaxies we consider is described in detail in Paper
II. It is composed of 68 galaxies in the Coma Cluster, 30 in Abell 85 and 34
in Abell 496; their cluster membership is confirmed by their redshifts. 

For all these galaxies, we have fitted the observed surface brightness
distribution (obtained by CCD imaging in the $V$ band) by means of the \Se\
integrated profile, i.e., using the growth curve fitting method (see Papers I
and II, for a detailed description and discussion of the fitting procedure).
As explained in our previous papers, S0 galaxies were excluded from our sample
because they fall out of the correlations between the S\'ersic parameters. To be
able to use all the data as a single set, the fitting parameters $a$ and
$\Sigma_0$ have been expressed in physical units, i.e., in kpc and
L$_\odot$/kpc$^2$ respectively\footnote{We have used $H_0=100$
km~s$^{-1}$Mpc$^{-1}$ and $q_0=1/2$}.

\begin{figure}[htb]
\centering
\mbox{\psfig{figure=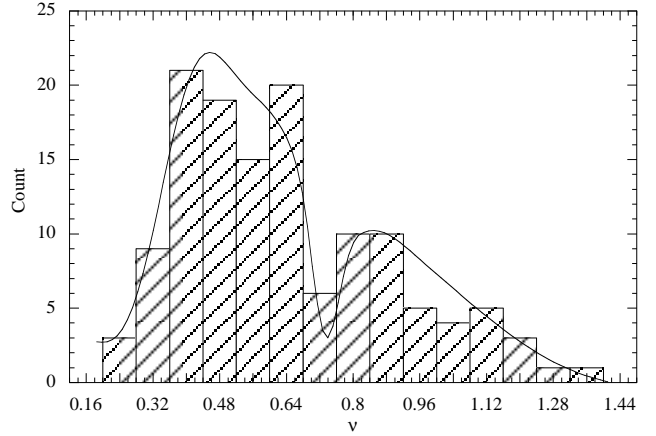,width=8.5cm}}
\caption[]{Histogram of the $\nu$ distribution
superimposed on the wavelet reconstruction of the density distribution.}
\label{fig:nudistribution}
\end{figure}

While in the de Vaucouleurs law a single value of $\nu$ is used ($\nu
= 1/4$), our data fitting reveals that this parameter can cover a
range of values, which are not centered on 0.25. We give in
Fig.~\ref{fig:nudistribution} the density distribution of $\nu$
obtained by wavelet reconstruction [see \citet{Fadda} for a
description of the method]. The distribution is bi-modal, with a first
maximum around $\nu=0.44$ and a second, less pronounced, one around
$\nu = 0.82$ (close to 1, the typical value for dwarf spheroidal
galaxies); notice that in the previously published $\nu$-histogram
(Paper I), this second maximum is already present, although not as
significantly. This is due to the fact that we are now working with a
larger sample, resulting from the combination of three separate data
sets. It is worth noting however that our sample is more than 97\%
complete only up to magnitude $V \sim 20$ and the giant ellipticals of
Coma were not included because of saturation of the CCD; besides, the
CCD fields do not cover the entire clusters, but only the central
parts (in particular for Abell 85 and 496). Therefore both the
position of the two peaks and their relative intensities may be
somewhat different when considering complete samples.

\subsection{Entropy and energy--mass relations}\label{subsec:entropydata}

As discussed in Sect.~\ref{subsec:remarques}, the mass, entropy and
potential energy of all the galaxies in our sample were computed under the
hypothesis that light traces mass.

The uniqueness of the specific entropy of Es has been discussed and
established in Paper I, at least at zeroth order. In Paper II we have
shown that, as a result of energy and mass exchange in merging
processes, $s_{0}$ weakly depends on the mass.  A constant value
for the mass-to-light ratio was assumed in these two papers.

\begin{figure}[htb]
\centering
\mbox{\psfig{figure=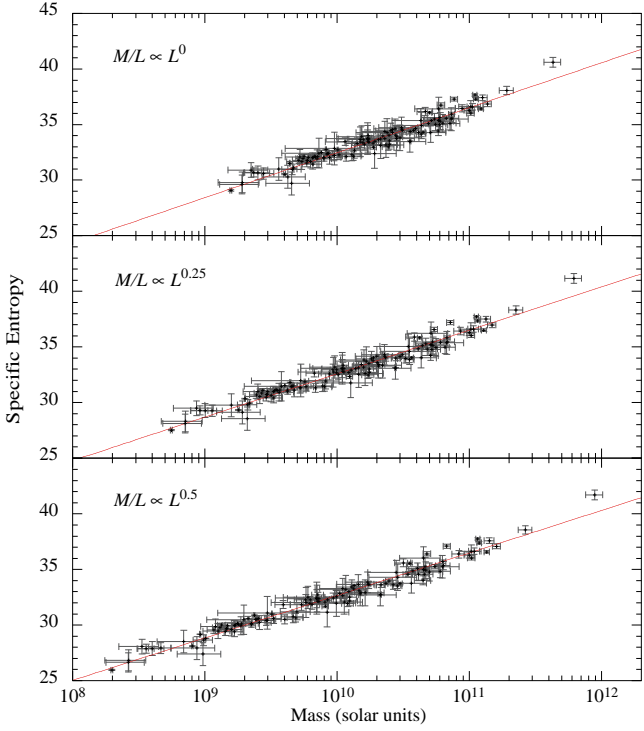,width=8.5cm}}
\caption[]{Specific entropy versus mass for three values of
$\alpha$. The bisector-OSL fits to our 132 galaxy sample are shown as
full lines.}
\label{fig:sversusL}
\end{figure}

We now calculate the specific entropy as a function of mass for three mass to
light relations: $M/L \propto L^0$ (constant mass to light ratio),
$L^{0.25}$ (from Fundamental Plane studies, e.g., \citet{Burstein}) and
$L^{0.5}$ (extreme case). The results are shown in Fig.~\ref{fig:sversusL} and
the best power law fits corresponding to Eqs.~(\ref{eq:tiltEntro}) and
(\ref{eq:U-M}) are given in Table~\ref{tbl:stats}.

\begin{figure}[htb]
\centering
\mbox{\psfig{figure=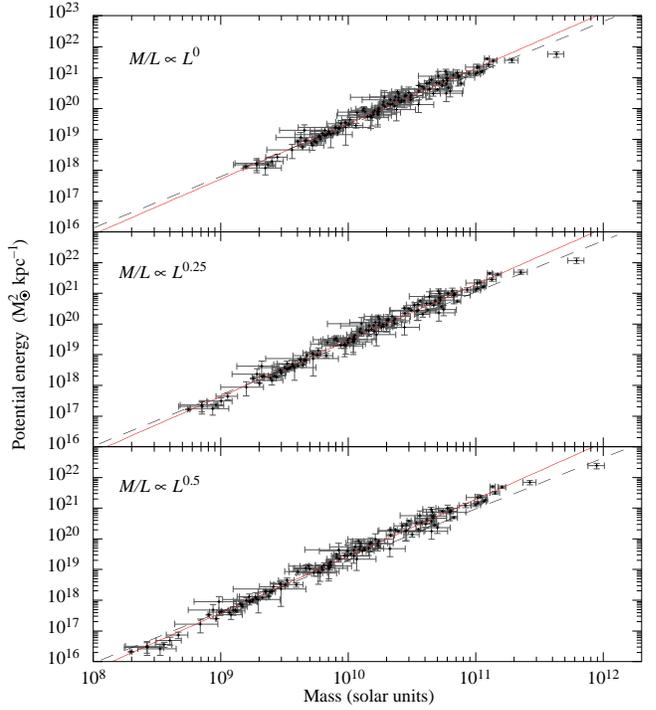,width=8.5cm}}
\caption[]{Potential energy (with $G=1$) versus mass for three values of
$\alpha$. The power-law fits (full lines) give slopes of 1.79, 1.83 and 1.85
respectively from top to bottom (see Table \ref{tbl:stats}. The theoretical 
relation, with an index of 5/3, is also shown (dashed lines).}
\label{fig:energieversusL}
\end{figure}

The potential energy is displayed in Fig.~\ref{fig:energieversusL} as a
function of total mass (adopting $G=1$). A bisector-OLS (Ordinary Least
Square) power law fit \citep{Feigelson} gives a indexes $\beta = 1.79$, 1.83
and 1.85 for $M/L \propto L^0, L^{0.25}$ and $ L^{0.5}$ respectively (see
Table~\ref{tbl:stats}). However, if we impose $\beta = 5/3$ [as predicted in
Sect.~\ref{subsec:scaling}, see Eq.~(\ref{eq:ouf})], the fit is still good
enough (within $2\sigma$, cf. Table~\ref{tbl:stats}) to confirm that
Eq.~(\ref{eq:econst}) is a good approximation. \cite{Fish} analysed a set of
24 Es, fitting their surface brightness profiles with the de Vaucouleurs law,
and found a power law index $\beta = 3/2$. The improved accuracy of both our
data and our fitting procedure (using the \Se\ profile), as well as our larger
sample, allow us to definitely exclude $\beta = 3/2$ (see also
\S~\ref{halos}).

%%%%%%%%%%%%%%%%%%%%%%%%%%%%%%%
\begin{table}[htb]
%\centering
\caption{Fitting results for the specific entropy and potential energy 
scaling relations. The slope of the mass--luminosity ratio is $\alpha$, 
$s_{0}$ and $\delta_{s}$ are the parameters of the entropy--mass relation; 
$e_{0}$ and $\beta$ are energy--mass parameters.}
\begin{tabular}{ccccc}
\hline
        $\alpha$   & $s_{0}$ & $\delta_{s}$ & $e_{0}$ & $\beta$ \\
\hline
    0.0~ & $-8.0\pm 1.2$ & $1.76\pm 0.06$ & $3.68\pm 0.03$ & $1.79\pm 0.06$ \\
    0.25 & $-6.5\pm 1.0$ & $1.70\pm 0.04$ & $2.74\pm 0.05$ & $1.83\pm 0.06$ \\
    0.50 & $-5.5\pm 0.8$ & $1.66\pm 0.03$ & $2.08\pm 0.06$ & $1.85\pm 0.06$ \\
           \hline
\end{tabular}

\begin{footnotesize}
Notice that the values of $s_{0}$ for $\alpha$ = 0 are different from the ones
shown in Paper II due to our different definition of the specific
entropy (cf. Sect.~\ref{sec:Sersic} and Table~\ref{tbl:formule2}).
\end{footnotesize}
\label{tbl:stats}
\end{table}
%%%%%%%%%%%%%%%%%%%%%%%%%%%%%%%%%%%%%%%%%%%%%%%%%%%%%%%%%%%%%

\subsection{Correlations between the S\'ersic profile parameters}%
\label{subsec:correlationsdata}

From a mathematical point of view the primary parameters $a$,
$\Sigma_0$ and $\nu$ entering in the definition of the \Se\ profile
are independent from each other. But as discussed in
Sect. \ref{sec:Sersic}, the entropy and energy--mass relations,
Eq.~(\ref{eq:PlanEntroTilt}) and (\ref{eq:PlanEnergMass}), imply the
existence of two-by-two relations between the three S\'ersic
parameters. From the observational point of view, we know that a,
$\Sigma_{0}$ and $ \nu$ are in fact correlated \citep{Young1,
Gerbal}. We now explain these correlations as originating from the
physical laws discussed in Sect. \ref{sec:deuxlois}.

\begin{figure*}
\centering
\mbox{\psfig{figure=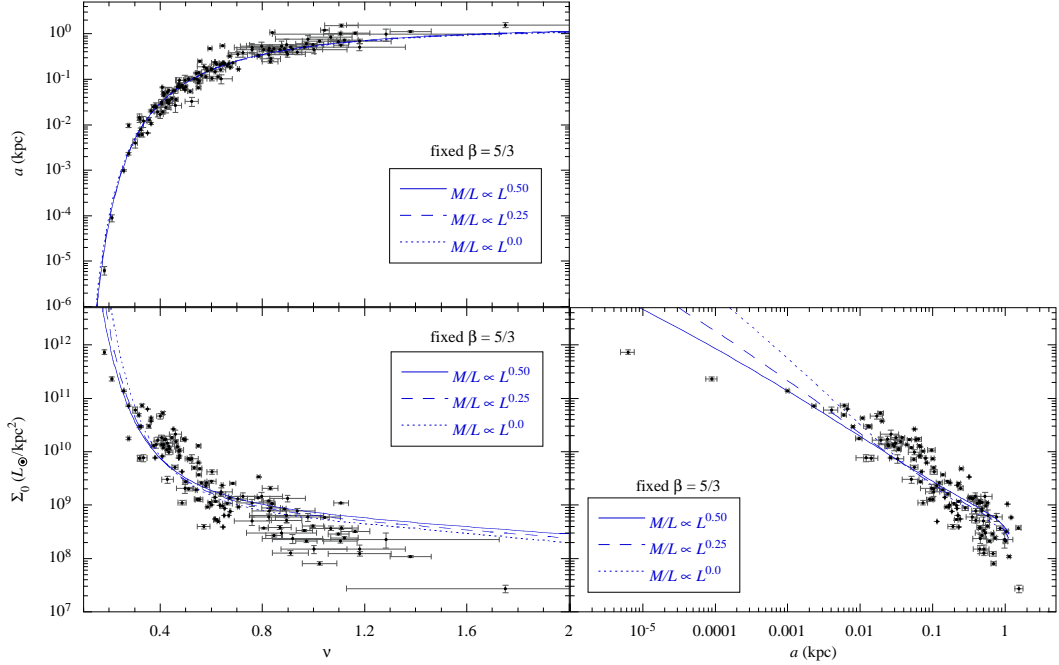,width=14cm}}
\caption[]{Pairwise correlation between the S\'ersic parameters [$a, \nu$],
[$\Sigma_{0}, \nu$] and [$\Sigma_{0}, a$] (counter clockwise from top left).
The lines superimposed in each panel are the correlations predicted
from the theoretical slope $\beta$=5/3 for $\alpha$=0.0, 0.25 and 0.50.}
\label{fig:a_nu_sig0_theoFin_a}
\end{figure*}

\begin{figure*}
\centering
\mbox{\psfig{figure=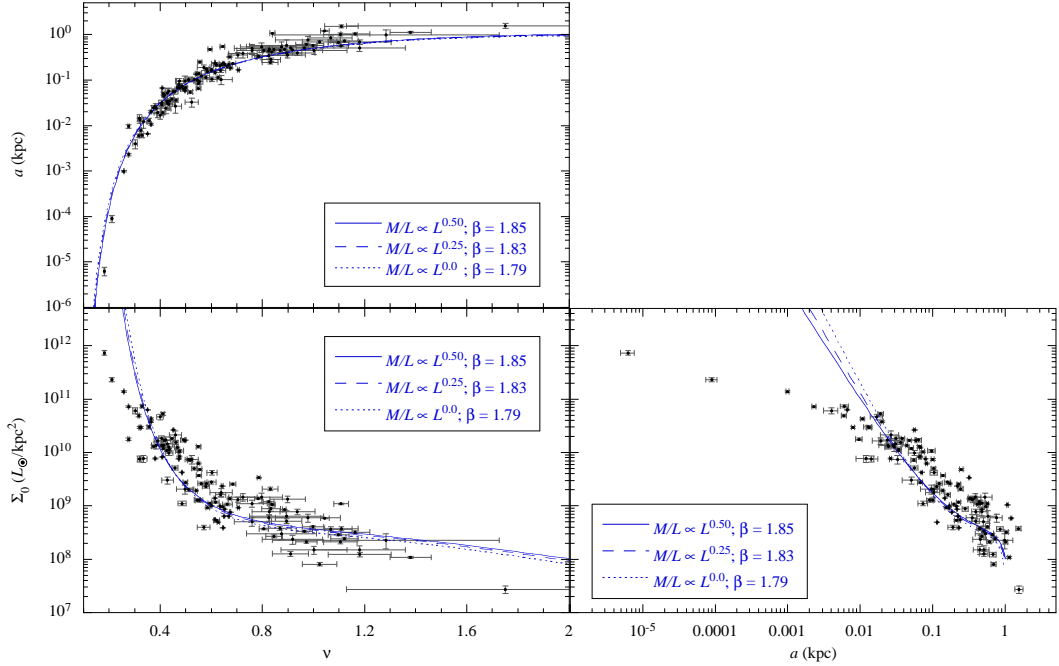,width=14cm}}
\caption[]{Pairwise correlation between the S\'ersic parameters [$a,
\nu$], [$\Sigma_{0}, \nu$] and [$\Sigma_{0}, a$] (counter clockwise
from top left).  The lines superimposed in each panel are the
correlations predicted from the Entropic Surface (with $\alpha$=0.0,
0.25 and 0.50) and the fitted $\beta$ values for the energy--mass scaling 
relations for these three values of $\alpha$ (see Table~\ref{tbl:stats}).}
\label{fig:a_nu_sig0_theoFin_b}
\end{figure*}

Figs.~\ref{fig:a_nu_sig0_theoFin_a} and \ref{fig:a_nu_sig0_theoFin_b}
show the [$a, \nu$], [$\Sigma_{0}, \nu$], and [a, $\Sigma_{0}$]
relations for our sample of galaxies. The corresponding theoretical
relations, computed for $\alpha$=0.0, 0.25, 0.50 are superimposed on
the data for two cases: the fixed value of $\beta$=5/3 (the
theoretical prediction as in Eq.~\ref{eq:relechelle}) and the values
of $\beta$ given in Table~\ref{tbl:stats}. The theoretical prediction
lies close to the data points, in particular for big galaxies (those
with the smallest $\nu$ for $\beta$=5/3 -- see section
\ref{subsec:discussions}).  We note that in both cases the effect of
changing the M/L ratios is small.

%We also plot a 3-D representation
%of these correlations (with $\alpha$=0 for simplicity) in the S\'ersic
%parameter space. 

%---------------------------------------------------------------------
%---------------------------------------------------------------------

%---------------------------------------------------------------------
%---------------------------------------------------------------------

\subsection{From theoretical to observed correlations}\label{sec:relations}

The well known photometric correlations which characterize elliptical
galaxies, are usually expressed by means of astrophysical quantities like
$R_\eff$, ${\langle \mu\rangle}_\eff$, and/or the absolute magnitude of the
galaxies. Since these quantities are combinations of the \textit{primary}
parameters defining the S\'ersic law, if these observed photometric correlations
do indeed contain information on the physics of the objects or on the
processes that drove their formation and evolution, one should, in principle,
be able to derive them from the ``theoretical'' relations,
Eqs.~(\ref{eq:entro}) and (\ref{eq:econst}). In doing so, we show that it is
indeed possible to give a physical interpretation to some of the observed
correlations and that they are not just artefacts of the definitions of the
parameters.

Before going through the comparison between theory and observation however, it 
is necessary to take into account the following points:
\begin{enumerate}
\item The correlations proposed in the literature \citep[e.g.,][]{Young0,
Binggeli} result from fitting procedures; the coefficients are just empirical.
Moreover, for the sake of simplicity, attempts are usually done to linearize the
correlations. On the contrary, the coefficients that we derive are the
consequence of physical laws, and the relations we have found are non-linear.

\item The intersection of the \SE\ and of the \EM\ surfaces is a line
that we call the ``\EEL''. The region of this line populated by
ellipticals is not coplanar, but can be considered as located in a
plane in a restricted range of values.

\end{enumerate}

\subsubsection{Luminosity--effective radius relation}\label{subsec:L-Reff}

From the intersection of the \SE\ and of the \EM\ surfaces,
Eqs.~(\ref{eq:lna}) and (\ref{eq:lnSigma0}) parametrized by $\nu$, we can
derive photometrical relations for Es galaxies using the formulae in 
Table~\ref{tbl:formule1}.

\begin{figure*}[htb]
\centering
\mbox{\psfig{figure=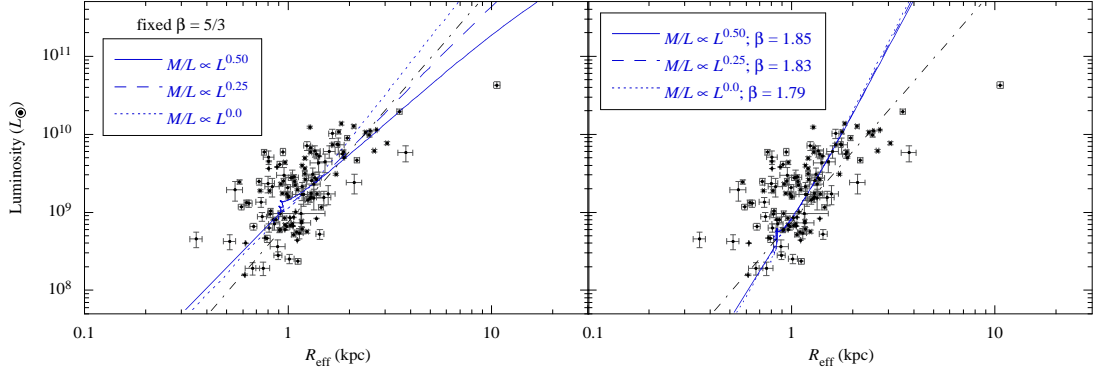,width=14.5cm}}

\caption[]{Luminosity versus effective radius for our 132 cluster ellipticals.
The theoretical relations obtained for a fixed $\beta$=5/3 (left) and for the
fitted values of $\beta$ (right) are superimposed for the three different
values of $\alpha$. The theoretical relation given by Eq.~(\ref{eq:L-reff}) is
superimposed as a dot-dashed line. The continuous, dashed and dotted lines are 
obtained using the intersection of the specific entropy--mass and energy--mass 
relations, Eqs.~(\ref{eq:PlanEntroTilt}) and (\ref{eq:PlanEnergMass}).
}
\label{fig:Lum-Reff}
\end{figure*}

In Fig.~\ref{fig:Lum-Reff} we plot
the total luminosity as a function of $R_\eff$ for our data.

%---------------------------------------------------------------------
%---------------------------------------------------------------------

Using the definition of the potential energy, relation~(\ref{eq:econst}) can be
written as:
\begin{equation} e_{0} = \frac{1}{3} [\ln M - 3\ln r_{g}] \, ,
\label{eq:ouf3}
\end{equation}
where $r_{g}$ is the gravitational radius and $M$ the mass. Assuming
that $R_{\eff}$ is proportional to $r_{g}$ (which is a very good
approximation for galaxies described by the \Se\ profile, see
Table~\ref{tbl:formule1}), we then have:
\begin{equation}
    \log L = 3 \log R_{\eff} + 3 \log k - \log\mathcal{F} +1.30288 e_0 \, ,
    \label{eq:L-reff}
\end{equation}
where $\mathcal{F}$ is the $M/L$ ratio and $k$ is the proportionality constant
between $R_{\eff}$ and $r_{g}$. This relation is drawn with the dot-dashed
line on Fig.~\ref{fig:Lum-Reff}. Notice that the theoretical
slope of the $L$--$R_\eff$ relation, in this case, depends only on the slope
of the scaling relation between mass and potential energy.

We can relax some simplifying hypotheses, proportionality between $R_{\eff}$
and $r_{g}$ and constant $M/L$ ratio and obtain a more general relation
between $L$ and $R_{\eff}$ -- the price to pay is a much more complicate
expression, instead of Eq.~(\ref{eq:L-reff}). The theoretical relations
obtained for a fixed $\beta$=5/3 (left) and for the fitted values of $\beta$
(right) are superimposed for the three different values of $\alpha$. Within
the error bars, the relations are compatible with the data. The different
curves vary with M/L when $\beta$ is fixed ($\beta$=5/3) but lead to similar
curves when $\beta$ is free.

Our data and theoretical prediction are steeper than the results found
in the literature. \cite{Binggeli0} derive $L \propto R_{\eff}^{2}$
for galaxies with $0.2 \la R_{\eff} \la 5$~kpc. For smaller galaxies,
however, the slope steeps and reaches our theoretical value of
3. \cite{Schombert} also finds that the $L$--$R_\eff$ relation is
shallower for the brightest galaxies; for normal ellipticals he finds
$L \propto R_{\eff}^{1.8}$. Even though these low values are not
incompatible with our data (within the error bars), it should be
stressed that previous $L$--$R_\eff$ studies were based on the de
Vaucouleurs profile, while we fit our galaxies with the \Se\ law. As
we will see below, different modelling (i.e., de Vaucouleurs or \Se\
laws) of Es may change the observed correlations among their global
parameters.

\subsubsection{Magnitude--mean surface brightness correlation}\label{subsec:M/mueff}

\begin{figure*}[htb]
\centering
\mbox{\psfig{figure=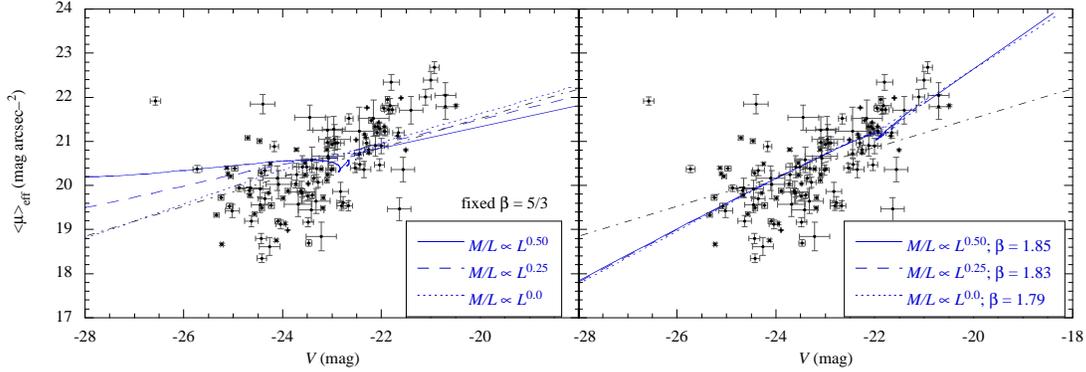,width=14.5cm}} 
\caption[]{Mean effective magnitude versus absolute magnitude. The results for
the different values of $\alpha$ are plotted for the theoretical prediction
$\beta$=5/3 (left) and for the fitted values of $\beta$ (right). The
theoretical relation given by Eq.~(\ref{eq:jerjen}) is superimposed as a
dot-dashed line. The continuous, dashed and dotted lines have the same meaning
as in Fig.~\ref{fig:Lum-Reff}.}
\label{fig:mue-mag}
\end{figure*}

Taking the definition of $\langle\mu\rangle_\eff$:
\begin{equation} \langle\mu\rangle_\eff = -2.5 \log L + 5
\log R_\eff + 2.5 \log(2 \pi) \, ,
\label{eq:mueff}
\end{equation}
we can combine it with the energy--mass scaling relation,
Eq.~(\ref{eq:ouf3}), eliminating the effective radius. The results for
the different values of $\alpha$ are plotted in
Fig.~\ref{fig:mue-mag}.  The theoretical relations obtained for a
fixed $\beta$=5/3 (left) and for the fitted values of $\beta$ (right)
are superimposed for the three different values of $\alpha$. The
different curves vary with M/L when $\beta$ is fixed ($\beta$=5/3) but
lead to similar curves when $\beta$ is free.

If we make the same assumptions as in
\S~\ref{subsec:L-Reff} above, we obtain a relation between
$\langle\mu\rangle_\eff$ and $L$, or equivalently, in terms of the absolute
magnitude ($\mathcal{M} \equiv -2.5\log L$):
\begin{equation}
    \langle\mu\rangle_\eff = \frac{1}{3} \mathcal{M} -5 \log k
     + \frac{5}{3} \log \mathcal{F} -2.1715 e_{0}  + \frac{5}{2} \log(2 \pi)
     \, ,
    \label{eq:jerjen}
\end{equation}
traced as a dot-dashed line in Fig.~\ref{fig:mue-mag}, where the variables
have the same meaning as in Eq.~(\ref{eq:L-reff}). Note that this relation
does not predict two regimes but a single one. This is indeed in agreement
with the observations when the data are analysed with the \Se\ law, as done by
\citet{Jerjen}, and as shown in Fig.~\ref{fig:mue-mag} for our galaxies.
Quantities like $R_\eff$ or $\langle\mu\rangle_\eff$ are different when
obtained through a \DV\ or a \Se\ fit \citep{Graham, Jerjen}: this probably
explains why \citet{Binggeli0}, who used the de Vaucouleurs profile, obtain a
broken power-law $\langle\mu\rangle_\eff$--$\mathcal{M}$ relation, while
\citet{Jerjen} and us, using the \Se\ profile, find a single power-law.

%---------------------------------------------------------------------------
%-------------------------------------------------------------------------------

\subsubsection{Photometric plane}\label{subsec:photometrie}

\cite{Khosroshahi} have fitted a set of infra-red profiles of E and spiral
bulges with the \Se\ law. They have shown that these galaxies lie on a
plane in the space defined by the set of coordinates $[\log R_{\eff}, \mu_{0},
\log \nu]$, with $\mu_{0}= -2.5 \log \Sigma_{0}$. Notice that these authors
use $n = 1/\nu$ in the \Se\ law. The equation they find for this plane is:
\begin{eqnarray}
(0.173\pm 0.025) \log R_\eff - (0.069\pm0.007) \mu_{0} = & & \nonumber \\
- \log \nu - (1.18\pm 0.05) \, . & &
\label{eq:khosro}
\end{eqnarray}

\begin{figure*}[htb]
\centering
\mbox{\psfig{figure=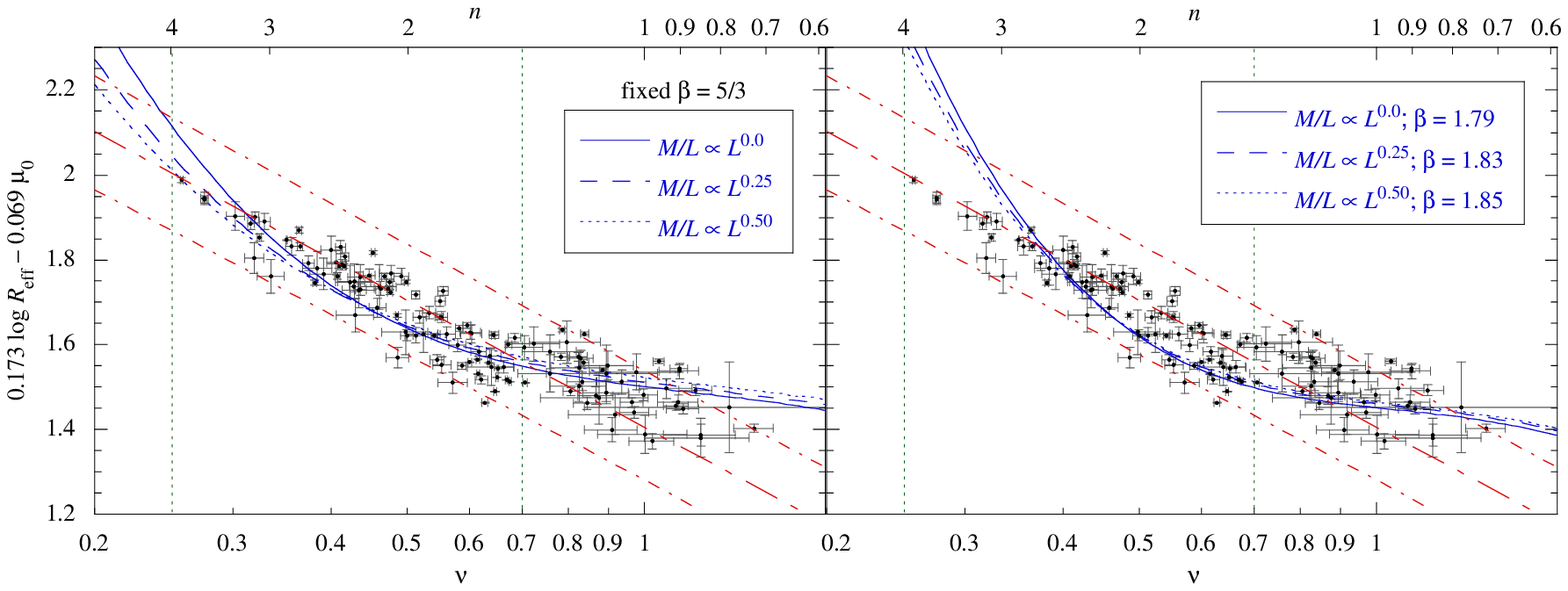,width=14.5cm}} 
\caption[]{The theoretical relation given by the generalized form of the
right-hand side of Eq.~\ref{eq:khosrocalcul} {\bf (solid, dashed and dotted lines)} 
is superimposed on the fit {\bf (long-short dashed line)} obtained by 
\citet{Khosroshahi} (right-hand side of
Eq.~\ref{eq:khosro}). The two vertical dotted lines correspond the range in
$\nu$ used by \citet{Khosroshahi}. The {\bf dash-dotted} lines correspond to 
8\% errors on the fit. The shape parameter, $n \equiv 1/\nu$, is given at the 
top of the figure to allow direct comparison with other authors.}
\label{fig:photo}
\end{figure*}

We will now derive a similar equation from a theoretical point of view. For
illustration purpose, we derive here the simplest case with constant mass to
light ratio ($\alpha = 0$) and $\beta = 5/3$. The general case is of course
similar but more cumbersome. Including the definition of $\mu_{0}$ and the
relation between $R_\eff$ and $a$ into Eqs.~(\ref{eq:PlanEntroTilt}) and
(\ref{eq:PlanEnergMass}), after some straightforward algebra, we obtain two
equations:
\begin{equation}
\log R_\eff - \frac{2}{15} \mu_{0} = \log R_\eff^{*}(\nu) 
 + \frac{2}{9} \log e \, \left(s_{0} - F_{2}(\nu) \right) \, ,
 \label{eq:transformation1}
\end{equation}
and
\begin{eqnarray}
 \log R_\eff + \frac{2}{5} \mu_{0} & = & 
            \log M_{2}^{*}(\nu) - 1.94 \log R_\eff^{*}(\nu) \nonumber \\
 & & - 1.303 e_{0} - 1.5113 \, .
 \label{eq:transformation2}
\end{eqnarray}
A linear combination between Eqs.~(\ref{eq:transformation1}) and
(\ref{eq:transformation2}) gives:
%......................................................................
\begin{equation}
    0.173 \log R_\eff - 0.069 \mu_{0} = K(\nu) \, ,
\label{eq:khosrocalcul}
\end{equation}
%......................................................................
with:
%......................................................................
%
\begin{eqnarray}
K(\nu) & = & -0.086125 \log M_{2}^{*}(\nu) + 
              0.4262 \log R_\eff^{*}(\nu)\nonumber \\
       &  & + 0.1122 e_{0} + 0.025 (s_{0} - F_{2}(\nu)) + 0.13016 \nonumber \, .
\end{eqnarray}
%
%......................................................................
Relation~(\ref{eq:khosrocalcul}) is obtained from theoretical relations;
on the other hand, relation~(\ref{eq:khosro}) is a fit to the observed
data. The only differences between these two relations are the two
right-hand sides, i.e. ($- \log \nu$) should be compared to
$K(\nu)$. Only the forms of the two functions are to be compared since
the constant depends on the units chosen. In Fig.~\ref{fig:photo}, we
call attention to the interval [0.25--0.7] in $\nu$, which
\cite{Khosroshahi} have used to define their photometrical plane. In
particular, their relation given by Eq.~(\ref{eq:khosro}) and ours
[Eq.~(\ref{eq:khosrocalcul})] agree within 8\% in this range. The good
agreement between the observed points of Khosroshahi et al.  and our
theoretical relation shows that the photometric plane may be
understood as a consequence of the two laws discussed in the present
paper. Note however that it is possible to define a ``plane'' only
because a limited range of values has been considered for $\nu$ (see
the beginning of \S~\ref{sec:relations}).

\section{Discussion and conclusions}\label{sec:discussion}

\subsection{Ellipticals as a single parameter family}\label{subsec:discussions}

The \Se\ profile is certainly not the \textit{ultimate} profile to reproduce
the surface brightness of ellipticals, from brighter galaxies to dwarf
ellipticals [see \cite{Jerjen}]. 
The \Se\ profile cannot apply to fine features such as boxy isophotes,
central black hole, etc., but it allows a fit of sufficient quality
for our aims, since we use it only to describe large scale properties
\citep{Caon, Graham, Prugniel, Marquez}. A finer description, which is
out of the scope of this paper, would certainly require another
parameter to reproduce the whole profile.

Although the use of the \Se\ profile may seem essential to our calculations,
it is nothing but an analytical expression which works well from the
theoretical point of view (essentially because it provides a shape factor) and
also observationally (it gives good fits). In fact any other function would be
equally appropriate -- provided it allows to calculate the entropy, energy,
mass, etc. This is \textit{not} the case for the de Vaucouleurs law
which is characterized by two parameters only, and for this reason does not
allow to define the \SE\ or \EM\ surfaces. Consequently, it is not possible to
define the \EEL\ using the \DV\ law.

Notice that the \DV\ law is a homologous law, i.e., all models are deduced
from a unique template by two scaling transformations. The introduction of a
shape parameter in the \Se\ law destroys -- mathematically speaking -- this
homology. However, since all galaxies fitted by a \Se\ law roughly have
the same specific entropy, we can eliminate one parameter ($\Sigma_{0}$, say)
by replacing it with by its expression as function of the two remaining
parameters, $\Sigma_{0}$ = $\Sigma_{0}(a, \nu)$. Therefore we have --
mathematically speaking -- only two independent parameters, like the \DV\ law.
This procedure, however, does not lead to any kind of homology, for it is not
possible to define a `template' from which any \Se\ distribution could be
derived only by scaling transformations.

Since all elliptical galaxies also obey the \EM\ relation,
Eq.~(\ref{eq:econst}), we can further eliminate another parameter ($a$, say)
with the result that a single parameter is sufficient to characterize a
galaxy, for instance $\nu$. In this case every galaxy is defined by its shape
parameter $\nu$: generalizing the \DV\ law with the introduction of $\nu$
allows to gather all ellipticals into a single large family, at least from the
photometric point of view.

The mass can then be rewritten formally as a function of $\nu$ only as:
\begin{equation}
    M(\nu)=\frac{2\pi a^{2}(\nu)}{\nu } \Sigma_{0}(\nu) \mathcal{F} \,
           \Gamma\left(\frac{2}{\nu}\right) \, ,
    \label{eq:bijection}
\end{equation}
where the equations for $a(\nu)$ and $\Sigma_{0}(\nu)$ are given in
Eqs.~(\ref{eq:correlationsanu}) and (\ref{eq:correlationsigmanu}). Moreover,
very large galaxies are endowed with small values of $\nu$ while dwarfs have
large values of $\nu$ \citep[e.g.,][see also Paper I; the equation
for $a(\nu)$ together with the definition of $R_\eff$ also implies this
result]{Caon}. Thus, the parameter $\nu$ plays the role of a concentration
factor. This allows to understand why the \DV\ law cannot fit with good
confidence these two extreme cases.

Equation (\ref{eq:bijection}) above is in fact a bijection $\nu
\Longleftrightarrow M$. During the processes of formation and relaxation,
physics (described by the two laws discussed above) are necessarily acting so
that at the end the above bijection is actually verified.

\subsection{Distance indicators}

The interesting possibility of using the correlation between $[a,\nu]$ as a
distance indicator has been proposed by \cite{Young1}. This paper has been
followed by controversies \citep{Binggeli, Young2}, essentially due, as we
understand it, to the following question: are distance indicators
sufficiently reliable to be usable for single galaxies, or can they only be
applied to clusters globally?

Since the correlations on which are based the indicators proposed by
\cite{Young1} are consequences of the two laws discussed in this paper, it
appears possible to improve their quality. This is because the predicted
relations are essentially \textit{non-linear}, while the observed correlations
have been tentatively fitted by phenomenological \textit{linear} laws, as we
already pointed out.

This question has already been mentioned in Paper I and, in view of
the predicted theoretical relations presented here, we therefore hope
to improve the distance indicator based on the shape profile of Es (to
be analysed in depth in a forthcoming paper for a dozen clusters with
redshifts between 0.06 and 0.37).

\subsection{Conclusions}\label{sec:conclu}

We have shown both from theoretical reasons and from observations, that
elliptical galaxies obey a scaling relation between the potential energy and
mass (luminosity). In previous papers, we had already shown that Es share the
same specific entropy (Papers I) with a logarithm dependence on the galactic
mass (that may be due to an evolutionary process, e.g., hierarchical mergers
Paper II).

These two relations, $U_{p}$--$M$ and $s$--$\ln M$, give an explanation to
several observed correlations that have been proposed in the past by various
authors, such as the correlations between the shape factor and a length scale,
the correlation between the absolute magnitude and the central brightness, and
the photometric plane. Therefore they constitute a theoretical background for
a number of physical properties of elliptical galaxies.

The fact that elliptical galaxies lie on a line in the three dimensional space
of the S\'ersic parameters implies that Es are indeed a one-parameter family.
This has important implications for cosmology and galaxy formation and
evolution models. Furthermore, the \textit{Energy--Entropy line} could be used
as a distance indicator.

\begin{acknowledgements}
We acknowledge interesting discussions with Gary
Mamon. I.M. acknowledges financial support from the Spanish Ministerio
de Educaci\'on y Cultura and the Instituto de Astrof\'{\i}sica de
Andaluc\'{\i}a (C.S.I.C.). This work is financed by DGICyT grants
PB93-0139, PB96-0921, the \textsc{usp/cofecub} and the
\textsc{cnp}q/\textsc{cnrs} and \textsc{c.s.i.c.}/\textsc{cnrs}
 bilateral cooperation agreements. G.B.L.N.
acknowledges financial support from the \textsc{fapesp}. B.L. is
supported by a Marie Curie Training Grant (category 20), under the TMR
Activity 3 of the European Community Program.
\end{acknowledgements}

%===========================================================================

%=============================================================================

\appendix

\section{Formulae}\label{sec:formules}

General formulae useful for definitions or calculations of various
quantities linked with our definition of the \Se\ profile are given in
Tables~\ref{tbl:formule2} and \ref{tbl:formule1}.

%%%%%%%%%%%%%%%%%%%%%%% formulaire 2 %%%%%%%%%%%%%%%%%%%

\begin{table*}[htbp]

\centering
\caption{Various formulae related to the specific entropy and the
energy--mass relation obtained for gravitational systems described by the
S\'ersic law. The 2-D and 3-D refer to the dependence on $\Sigma_{0}$ and
$\rho_{0}$, respectively. The quantities $M^{*}, M_{3}^{*}$ and $r_{g}^{*}$ are
defined in Table~\ref{tbl:formule1}. Note the different normalization used for 
defining the specific entropy from Papers I and II.}
\begin{tabular}{|l|l|}
\hline
      Quantity & S\'ersic expression \\
\hline
\textbf{Specific entropy} & $s = - \int \diff^{3} r\, \diff^{3} v \,
f^*\, \ln f^* \, ; \ \int \diff^{3} r\, \diff^{3} v f^* = 1 $\\
&\\
2-D expression  & 
   $s =\frac{3}{2}\ln (\mathcal{F}(L) \Sigma_{0}) + \frac{9}{2}\ln a + F_{2}(\nu)\, ;$ \\
           & $F_{2}(\nu) \simeq -0.795\ln(\nu) -\frac{1.34}{\nu} +
           3.85\nu^{-1.29} + \ln \Gamma\left[\frac{2}{\nu}\right] - 0.822 $\\
&\\
3-D expression  & $s = \frac{3}{2} \ln (\mathcal{F}(L) j_{0}) + 6 \ln a + F_{3}\, ;$\\
                & $F_{3}(\nu) = \displaystyle{F_{2}(\nu) +\frac{3}{2}
                  \ln[2\frac{\Gamma([3-p]/\nu)}{\Gamma(2/\nu)}]}$\\
&\\
\textbf{Potential Energy } &\\
2-D expression & $|U| = G \, a^3 (2 \pi \mathcal{F} \Sigma_{0} )^2 \times
                  G_{2}^{*}(\nu) \, ;$\\
 &  $\displaystyle{G_{2}^{*}(\nu) \simeq \left( \frac{\Gamma[2/\nu]}{\nu 
     \Gamma\left[(3-p)/\nu\right]} \right)^2 \, \exp \left( 1.7958 
     \nu^{-3/2} + 2.6374 \sqrt{\nu} -5.4843 \right) } $\\
&\\

\textbf{Specific entropy--Mass} & $s = s_{0} + \delta_{s} \ln M$\\

\textbf{relation} & \\

&\\
\textbf{Energy--Mass} & $U = k_{U} M^{\beta}$ \\
\textbf{relation} & $e_0 = \ln U - \beta \ln M$\\
&\\
2-D expression & $3e_0 = \ln\Sigma_{0} -\ln a + \ln M_2^*(\nu) -
   3 \ln r_g^{*}(\nu) + 3\ln G$ \\
&\\
3-D expression & $3e_0 = \ln \rho_{0} + \ln M_{3}^*(\nu) - 3\ln r_{g}^{*}(\nu)
                         +3 \ln G$ \\
\hline
\end{tabular}
\label{tbl:formule2}
\end{table*}
%%%%%%%%%%%%%%%%%%%%%%%%%%%%%%%

%%%%%%%%%%%%%%%%%%%%%%% formulaire %%%%%%%%%%%%%%%%%%%

\begin{table*}[htbp]
\centering
\caption{General formulae used in this paper. 2-D and 3-D quantities are marked
with subscript ``2'' and ``3'', respectively. Quantities marked with a ``*''
are adimensional. Some of these expressions were first obtained in Paper I and 
are presented here for completeness.}
\begin{tabular}{|l|l|l|}
\hline
     Quantity & Theoretical Expression & Analytical Approximation  \\
\hline
\textbf{S\'ersic law }& &\\
2-D & $\Sigma(R)=\Sigma _{0}\exp (-(R/a)^{\nu }) $ &\\

    & $L_{2}(R) = a^{2} \Sigma_{0}  
      \displaystyle{\frac{2\pi}{\nu} \gamma \left[\frac{2}{\nu}, 
      \left(\frac{R}{a} \right)^{\nu} \right]}$ &\\[8pt]

3-D   & & $j(r) = j_{0}\left( \frac{r}{a} \right)^{-p}
                \exp\left[-(r/a)^{\nu} \right]$ \\
      & & $ p\simeq 1.0-0.6097\nu + 0.054635\nu^{2}$  \\

      & & $L_{3}(r) = a^{3} j_{0} 
      \displaystyle{\frac{4 \pi}{\nu} \gamma \left[\frac{3-p}{\nu}, 
      \left(\frac{R}{a} \right)^{\nu} \right]}$ \\[8pt]

\textbf{Mass profile}& $M/L \equiv \mathcal{F}(L) = k_{F} L^{\alpha}
      \, ; \ \mathcal{F}(R) = \mbox{constant}$ &\\
      
2-D  & $M(R) = \mathcal{F}(L) L(R)$ & \\
3-D  & & $M(r) = \mathcal{F}(L) L_{3}(r)$ \\[8pt]

\textbf{Total Mass} & $M = \mathcal{F}(L) L = k_{F} L^{\alpha + 1}$ &\\

2-D  & $M = k_{F} \left(a^{2} \Sigma_{0} L_{2}^{*} \right)^{\alpha+1} \, ;
      \ L_{2}^{*} = \displaystyle{\frac{2 \pi}{\nu} 
      \Gamma \left[\frac{2}{\nu} \right]}$ & \\

3-D  & & $M = k_{F} \left(a^{3} j_{0} L_{3}^{*} \right)^{\alpha+1} \, ;
     \ L_{3}^{*} = \displaystyle{\frac{4 \pi}{\nu} 
     \Gamma \left[\frac{3-p}{\nu} \right]}$ \\[8pt]

Total ``magnitude" &  $\mathcal{M} = -2.5\log L$ & \\
     &  $ ~\quad = -2.5\log \Sigma_{0} - 5\log a + m^{*}$
        & $m^{*} \simeq -0.30413\nu -1.70786\nu ^{-1.44265}$ \\[6pt]

Mean $\mu$ inside $R_\eff$ & 
              $\langle\mu\rangle_\eff = -2.5\log \Sigma_{0} + m^{*} + $ &\\
           &  $ \qquad\quad + 5\log(R_\eff^*) + 2.5\log 2\pi$  &\\
& &\\
\textbf{Radii} & $R_{i} = a \times R_{i}^*$  &\\

Effec. Radius 3-D&
$\gamma[(3-p)/\nu, R_{\rm 3, eff}^*] = \frac{1}{2}\Gamma[(3-p)/\nu]$
              &$\displaystyle{\ln R_{\rm 3, eff}^* \simeq
              \frac{0.72701 -0.9877 \ln \nu}{\nu} +0.07021} $ \\[5pt]
Effec. Radius 2-D&
$\gamma(2/\nu, R_\eff^*) = \frac{1}{2}\Gamma(2/\nu)$
          &$\displaystyle{\ln R_\eff^* \simeq 
             \frac{0.70348 -0.99625 \ln \nu}{\nu} -0.18722} $ \\[5pt]
Gravit. Radius &
$|U| \equiv G\, M^{2}/r_{g}$ &  \\
      & $r^{*}_{g} = \left[\Gamma(2/\nu)/\nu\right]^{2} / 
      G_{2}^{*}(\nu) $&  $\displaystyle{\ln r_{g}^* \simeq
                \frac{0.82032 -0.92446 \ln \nu}{\nu} +0.84543}$ \\[5pt]
		  
$r_{g}$ versus $R_\eff$&  &$\ln r_{g}^* \simeq 1.16 + 0.98 \ln R_\eff^* $\\

$\Sigma_{0} \Leftrightarrow j_{0}$&\begin{tabular}{l}
    $\displaystyle{a\times j_{0} = \Sigma_{0}\frac{L_{2}^*}{L_{3}^*}}$\\
    $\displaystyle{a\times j_{0} = \Sigma_{0}\times
       \frac{\Gamma(2/\nu)}{2\, \Gamma([3-p]/\nu)}}$
    \end{tabular}
   & $a \, j_{0} \simeq \Sigma_{0} \times $\\[-10pt]
   && $\times (0.076685 + 0.3253\nu-0.041245\nu^2)$ \\

\hline
\end{tabular}
\label{tbl:formule1}
\end{table*}

The relation between the specific entropy and the mass, Eq.~(\ref{eq:entro})
can expressed as a surface in the S\'ersic parameters space with the help of 
the mass to light ratio, Eq.~(\ref{eq:M/Lfonction}):
\begin{eqnarray}
 - \alpha'  \delta_{s}' \ln \Sigma_{0}  & = & \delta_{s}' \ln k_{F} + s_{0} +
     \nonumber  \\
     & + & \displaystyle{\left(\alpha' \delta_{s}'+ \frac{3}{2}\right)
              \ln \left[ \frac{2 \pi\Gamma(2/\nu)}{\nu}\right]} - F_{2}^{*}(\nu) +
     \nonumber  \\
     & + & \left(2 \alpha' \delta_{s}' - \frac{3}{2}\right) \ln a \, ,
    \label{eq:PlanEntroTilt}
\end{eqnarray}
where $\alpha' \equiv \alpha + 1$, $\delta_{s}' \equiv \delta_{s} - 3/2$ and 
the function $F_{2}^{*}(\nu)$ is defined in Table~\ref{tbl:formule1}.

The scaling relation between the potential energy and the mass gives 
another surface in the S\'ersic parameters space:
\begin{eqnarray}
    -\alpha' \beta' \ln \Sigma_{0} & = & \beta' \ln k_{F} + \ln k_{U} - \ln(4 \pi^2 G) +
    \nonumber \\
     & + & \displaystyle{ (\alpha' \beta' + 2) \ln \left[\frac{2 \pi 
                          \Gamma(2/\nu)}{\nu}\right] -  \ln[G_{2}^{*}(\nu)] +}
    \nonumber \\
     & + & [2(\alpha' \beta' + 2) -3] \ln a \, ,
    \label{eq:PlanEnergMass}
\end{eqnarray}
where $\alpha' \equiv \alpha + 1$, $\beta' \equiv \beta - 2$ and 
the function $G_{2}^{*}(\nu)$ is defined in Table~\ref{tbl:formule1}.

The intersection of the above two surfaces yields the locus where 
elliptical galaxies are found in the S\'ersic parameter space. Eliminating 
$\Sigma_{0}$ from Eq.~(\ref{eq:PlanEntroTilt}) or Eq.~(\ref{eq:PlanEnergMass}) 
we get:
\begin{eqnarray}
 (\frac{3}{2} \beta' + \delta_{s}') \ln a & = & \beta' [s_{0} -  F_{2}^{*}(\nu)] +
\nonumber \\
    & + & \displaystyle{(\frac{3}{2} \beta'- 2 
      \delta_{s}') \ln\left[\frac{2 \pi \Gamma(2/\nu)}{\nu}\right]} 
\nonumber \\
   & + & \delta_{s}' \ln[G 4 \pi^2 \, G_{2}^{*}(\nu) / k_{U}] \, .
    \label{eq:lna}
\end{eqnarray}
Note that $\ln a$ \textit{versus} $\nu$ depends on the mass to light ratio
only indirectly via $\delta_{s}$, $s_{0}$, $\beta$ and $e_{0}$. Likewise,
eliminating $\ln a$ from the surface equations we obtain:
\begin{eqnarray}
 (\frac{3}{2} \beta' + \delta_{s}') \alpha' \ln \Sigma_{0} & = & (2\alpha' 
 \beta'+1) (F_{2}^{*}[\nu] - s_{0})  \nonumber  \\ 
     & - & 3 \left([\frac{3}{2} \beta' - \delta_{s}'] \alpha' + \frac{3}{2} 
     \right)
     \ln\left[\frac{2 \pi \Gamma(2/\nu)}{\nu}\right]  \nonumber  \\ 
     & + & \left(\frac{3}{2} - 2 \alpha' \delta_{s}' \right) \ln[G 4 \pi^{2} \,
     G_{2}^{*}(\nu)/k_{U}] \nonumber \\
     & - & \left(\frac{3}{2} \beta' + \delta_{s}'\right) \ln k_{F} \, .
    \label{eq:lnSigma0}
\end{eqnarray}

The relation between $a$ and $\Sigma_{0}$ is obtained in parametric form
combining the above two equations and using $\nu$ as a parameter.
% ================================================================

\end{document}